\documentclass[12pt,nofootinbib]{revtex4}
\usepackage{graphicx}\usepackage{ifthen}\usepackage{bm}
\usepackage{amssymb}\usepackage{amsmath}\usepackage{latexsym}
\newcommand{\beq}[1]{\begin{equation}\label{#1}}
\newcommand{\eeq}{\end{equation}}
\newcommand{\ba}[1]{\begin{eqnarray} \label{#1}}
\newcommand{\ea}{\end{eqnarray}}

\newcommand{\rf}[1]{(\ref{#1})}

\newcommand{\gsim}{\mathrel{\vcenter{\hbox{$>$}\nointerlineskip\hbox{$\sim$}}}}
\newcommand{\half}{\frac{1}{2}}

\newcommand{\eps}{\epsilon}

\newcommand{\delfoam}{\delta_{\rm Foam}}
\newcommand{\etwenty}{10^{20}\,{\rm eV}}
\newcommand{\acrit}{a_{\rm crit}}

\newcommand{\Eth}{E_{\rm th}}
\newcommand{\Ecl}{E_{\rm class}}
\newcommand{\Eplus}{E_{\rm th}^+}
\newcommand{\Eminus}{E_{\rm th}^-}
\newcommand{\Ethmax}{E_{\rm th}^{\rm max}}

\begin{document}
\title{Space-Time Foam and Cosmic-Ray Interactions}

\author{Marcin Jankiewicz}
\email{m.jankiewicz@vanderbilt.edu}

\author{Roman V. Buniy}
\email{roman.buniy@vanderbilt.edu}

\author{Thomas W. Kephart}
\email{kephartt@ctrvax.vanderbilt.edu}

\author{Thomas J. Weiler}
\email{tom.weiler@vanderbilt.edu}

\affiliation{Department of Physics and Astronomy, 
Vanderbilt University, Nashville, Tennessee 37235}

\begin{abstract}
It has been proposed that propagation of cosmic rays at extreme-energy 
may be sensitive to Lorentz-violating metric fluctuations (``foam'').
We investigate the changes in interaction thresholds for cosmic-rays 
and gamma-rays interacting on the CMB and IR backgrounds, for a 
class of stochastic models of spacetime foam.  The strength of the foam
is characterized by the factor $(E/M_P)^a$, 
where $a$ is a phenomenological suppression parameter.
We find that there exists a critical value of $a$ (dependent on the 
particular reaction: $\acrit\sim 3$ for cosmic-rays, $\sim 1$ for gamma-rays), 
below which the threshold energy can only be lowered,
and above which the threshold energy may be raised, but at most by a factor of 
two.  Thus, it does not appear possible in this class of models to extend 
cosmic-ray spectra significantly beyond their classical absorption energies. 
However, the lower thresholds resulting from foam may have signatures
in the cosmic-ray spectrum.  
In the context of this foam model, we find that 
cosmic-ray energies cannot exceed the fundamental 
Planck scale, and so set a lower bound of $10^8$~TeV 
for the scale of gravity. 
We also find that suppression of $p\rightarrow p\pi^{0}$ 
and $\gamma\rightarrow e^{-}e^{+}$
``decays'' favors values $a\gsim \acrit$.
Finally, we comment on the apparent non-conservation of particle 
energy-momentum, and speculate on its re-emergence as dark energy
in the foamy vacuum.

\end{abstract}

\pacs{    \hspace{4cm} VAND-TH-03/XX}

\maketitle

%
%
\newpage
%
%
\section{Introduction: metric fluctuations as space-time foam}
\label{1.1}
The space-time metric tensor
$g_{\mu\nu}$ becomes a dynamical variable when gravity is quantized,
and space-time foam is the quantum mechanical uncertainty 
${\delta}g_{\mu\nu}$ in this variable.
We investigate the possibility that the foam has phenomenological consequences.

We work within the framework of 
foam models having a parameterization given by 
\beq{0}
{\delta}g_{\mu\nu}\,{\geq}\,
\left(\frac{l_{P}}{l}\right)^{a}{\sim}\left(\frac{t_{P}}{t}\right)^{a}\,\,,
\eeq
where $l_{P}\equiv\sqrt{\frac{{\hbar}\,G}{c^{3}}}$ is the Planck length
and $t_{P}\equiv\frac{l_{P}}{c}$ is the Planck time 
in a four-dimensional theory.
In theories with more than four dimensions, 
these scales could be larger, in fact, much larger, than the Planck scales.
The parameter ``$a$'' depends on the foam model \cite{AC1,Ng1}.
The covariance implicit in the fluctuation variable 
${\delta}g_{\mu\nu}$ puts space and time uncertainties on an equal 
basis, in contrast to the situation in non-relativistic quantum mechanics
where the length-momentum relation arises from operator commutators,
and the energy-time relation from a less-compelling argument 
using the Fourier decomposition.
The uncertainty relations given in Eq.~\rf{0}, and those that follow below,
are sometimes called ``Generalized Uncertainty Principles'' (GUP).
We will use this name.

From (\ref{0}) and the fact that $\delta l^{2}=l^{2}\delta g$, one
obtains the uncertainties for length and time:
\beq{0a}
\delta l\,{\geq}\,\frac{l}{2}\,\left(\frac{l_{P}}{l}\right)^{a} \,,
\quad\quad{\rm and\ }
\delta t\,{\geq}\,\frac{t}{2}\,\left(\frac{t_{P}}{t}\right)^{a}
\eeq
The quantum mechanical relations between length and momentum,
and time and energy, then lead to the equivalent expressions for the 
uncertainty:
\beq{1}
{\delta}E\,{\geq}\,\frac{E}{2}\,\left(\frac{E}{M_{P}\,c^2}\right)^{a}\,\,,
\quad\quad{\rm and\ }
{\delta}p\,{\geq}\,\frac{p}{2}\,\left(\frac{p}{M_{P}\,c}\right)^{a}\,\,,
\eeq
where $M_{P}\sim10^{28}eV$ denotes Planck mass. 
From here on we drop explicit mention of powers of $c$.
These uncertain lengths, energies, etc., 
make the lengths of four-vectors uncertain,
and so break Special Relativity.
Accordingly, we must single out a frame in which 
these uncertainties are defined. 
It is common to assume that the special frame is the cosmic rest frame, 
in which the cosmic microwave background (CMB) is isotropic.  We adopt this assumption.

The  uncertainty relations defined above can effect 
violations of Lorentz invariance (LIV) 
even in the weak-field, flat-space limit.
Two of the most studied cases are (i) modification of the 
energy-momentum conservation equations, $\Delta P^{\mu}=0$, 
or (ii) modification of the energy-momentum dispersion relation,
$p^{\mu}\,p_{\mu}=m^2$.
Of course, LIV may also appear in both (i) and (ii) simultaneously.
In this work, we study LIV of the energy-momentum conservation law 
at the scales defining the GUP, 
but maintain the usual dispersion
relation $E_{i}^{2}=p_{i}^{2}+m_{i}^{2}$.


\section{Modified threshold energy for UHECR absorption}
Following many others \cite{AC1}-\cite{Recent},
we investigate the role that LIV-kinematics may play on 
ultra-high energy cosmic rays (UHECRs).
The UHECRs are gamma-rays at $E\sim$~tens of TeV, and 
nucleons at $E\sim\etwenty$.
Standard particle and astrophysical arguments lead one to expect 
TeV gamma-rays and $\etwenty$ nucleons to be just above 
their respective thresholds 
for absorption on cosmic radiation background fields.
The annihilation reaction for TeV~gamma-rays is
$\gamma+\gamma_{IRB}{\rightarrow}e^{+}+e^{-}$,
where $\gamma_{IRB}$ denotes a cosmic infrared background photon;
the cms energy threshold is $\sqrt{s}=2m_e$.
The energy-loss reaction for $\etwenty$~nucleons is 
dominated by
$N+\gamma_{CMB}{\rightarrow}\Delta\rightarrow N'+\pi$
near threshold,
where $\gamma_{CMB}$ denotes a photon in the CMB;
the cms threshold energy is $\sqrt{s}=m_\Delta$.

The generalized uncertainties can in principle 
raise or lower the energy thresholds of these reactions.
Raising an annihilation or energy-loss 
threshold presents the possibility of 
extending the CR spectrum beyond expected cutoff energies.
There have been suggestions that such extended spectra do exist
for both gamma-rays \cite{gam-anom} and for nucleons \cite{AGASA}.
Predictably, there have also been suggestions that LIV is the origin of
the anomalous spectra \cite{Ng1}.

Consider the general $2\rightarrow 2$ scattering 
reaction $a+b{\rightarrow}c+d$.  Let
$m_{a},m_{b},m_{c}$ and $m_{d}$ label the particle masses,
and $E_{a},E_{b},E_{c}$ and $E_{d}$ the particle energies.
The unmodified dispersion relation reads
\beq{dispreln}
E_{i}=p_{i}+\frac{m_{i}^{2}}{2p_{i}}
	+{\cal O}\left(\frac{m_{i}^{4}}{p^3_{i}}\right)\,,
		\quad i=a,b,c,d\,\,.
\eeq
However, the energy and momentum conservation laws 
{\sl including GUP fluctuations} become
\begin{eqnarray}\label{2}
E_{a}+\delta{E}_{a}+E_{b}&=&E_{c}+\delta{E}_{c}+E_{d}+\delta{E}_{d}\,\,\,,\cr
p_{a}+\delta{p}_{a}+p_{b}&=&p_{c}+\delta{p}_{c}+p_{d}+\delta{p}_{d}\,\,.
\end{eqnarray}
According to Eq.~(\ref{1}),
the energy and momentum fluctuations of the cosmic background photons,
${\delta}E_{b}$ and ${\delta}p_{b}$, 
are very small since $E_{b}$ and $p_{b}$ are so small 
($\sim$~meV for the CMB and a few tens of meV for the far-IRB).
We have set them to zero.
The magnitudes of the other uncertainties can be significant.

Subtracting the second of Eqs.~(\ref{2}) from the first,
inserting \rf{dispreln} for the differences $E-p$, 
and realizing 
\footnote
{we choose to work in a frame where CR momentum 
is $\vec{p}_{CR}=p_{CR}(0,0,\hat{z})$, and
momentum of background photon is $\vec{p}_{B}=p_{B}(0,0,-\hat{z})$}
that $p_b=-E_b$, 
one gets for the energy of the background photon,
\beq{2a}
E_{b}=E_b^0 +\delfoam\,,
\eeq
where
\beq{Ebclass}
E_b^0=\frac{1}{4}\,
\left(\frac{m_{c}^{2}}{p_{c}}+\frac{m_{d}^{2}}{p_{d}}-\frac{m_{a}^{2}}{p_{a}}\right)
\eeq
is the classical value, and
\beq{Ebfoam}
\delfoam=\delta_{c}+\delta_{d}-\delta_{a}\,,
	\quad{\rm with\ }\delta_j=\half\,(\delta{E_j}-\delta{p_j})\,,
\eeq
is the contribution from fluctuating foam.
To first order in $m_j^2$, the $p_j$'s in Eq.~\rf{Ebclass} 
can be replaced by $E_j$'s.
We remark at this point that $\delfoam$ is sourced by 
$\delta E$ and/or $\delta p$.
Therefore, the origin of $\delfoam$ is open to a broad interpretation.
We discuss this a bit more in section \rf{sec:discuss}.
Also, since $\delta p$ is a component of a three-vector,
whereas $\delta E$ is not,
the relative sign in the combination $\delta E-\delta p$ appearing in 
the definition of $\delfoam$ is not meaningful.

Next, consider the reaction at threshold, $\sqrt{s}=m_{c}+m_{d}$.
In terms of the boost factor $\gamma$ between the 
center of mass frame and the lab frame,
one has 
$E_{\rm tot}^{LAB}=\gamma (m_{c}+m_{d})=E_a+{\cal O}(E_b)$,
$E_{c}=\gamma m_{c}$, 
and $E_{d}=\gamma m_{d}$,
i.e.,
$\gamma=\frac{E_{c}}{m_{c}}=\frac{E_{d}}{m_{d}}=\frac{E_{a}}{m_{c}+m_{d}}$.
These equalities allow the elimination of $E_c$ and $E_d$ in 
Eq.~\rf{Ebclass} in terms of $E_a$, 
which we write as $\Eth$ to remind ourselves that
the kinematics are being calculated at threshold energies.
The result is that Eq.~\rf{2a} becomes
\beq{2b}
4\,E_{b}\,\Eth=(m_{c}+m_{d})^{2}-m_{a}^{2}+\Eth\,\delfoam\,.
\eeq
Solving this equation for $\Eth$ then gives the modified threshold energy
for the reaction.
Of course, some model for the fluctuations must be introduced.


%
References \cite{Ng1,Ng2,AC1} argued for the following 
form of the fluctuations:
\beq{3}
\delta_j\equiv\half\,(\delta{E}_j-\delta{p}_j)=
	-\frac{\eps}{4}\,p_j\,\left(\frac{p_j}{M_{P}}\right)^a
	\approx-\frac{\eps}{4}\,E_j\,\left(\frac{E_j}{M_P}\right)^a\,.
\eeq
Different choices for $a$ and $\eps$ 
parametrize different space-time foam models. 
With dimensionful mass-energy factors explicitly shown,
$\eps$ is expected to be a number roughly of order one.
The exponent $a$ is assumed to be positive such that fluctuations are 
suppressed below the Planck scale.
Possibilities other than the 
particular form for the fluctuations given in Eq.~\rf{1}
are certainly possible.
For example, one may choose to parameterize the fluctuation 
$\delfoam$ in Eq.~\rf{2b} directly, with a parameterization of one's choosing.
However, we will stay with the form given above.

As we have seen, the individual $E_j$'s are linearly related to each other 
by mass ratios.  Along with Eq.~\rf{3}, this means that the $\delta_j$'s
are also (nonlinearly) related to each other by mass ratios.
This makes the result of inserting 
Eqs.~\rf{3} for each $j=a,\,,b\,,c$ into Eq.\ \rf{2b} fairly simple.  
The result is a general and manageable equation for the 
reaction threshold energy, 
incorporating the correction from space-time foam:
\beq{4}
4\,E_{b}\,\Eth=4\,E_b\,\Ecl +\eps\,\Eth^2\,\left(\frac{\Eth}{M_P}\right)^{a}\,
	\left[1-\frac{m_{c}^{1+a}+m_{d}^{1+a}}{(m_{c}+m_{d})^{1+a}}\right]\,,	
\eeq
with $\Ecl=\frac{(m_c+m_d)^2-m_a^2}{4\,E_b}$ being the energy of the 
threshold when special relativity is not violated.
When $\eps=0$, 
the classical threshold $\Eth=\Ecl$ of course obtains.
When $a=0$, the classical threshold $\Eth=\Ecl$ also obtains,
for any value of $\eps$.

Before examining specific models for the meaning of $a$ and $\eps$,
we can extract from Eq.~\rf{4} the number of real positive roots of $\Eth$.
These roots are the candidate solutions for the modified threshold.
Imagine plotting the LHS and RHS of the equation versus $\Eth$.
The LHS of \rf{4} rises linearly in $\Eth$ from zero,
with a slope of $4\,E_b$. 
The RHS rises (falls) for positive (negative) $\eps$ 
at a higher power of $\Eth$, 
from a positive intercept of $(m_c+m_d)^2-m_a^2$.
For negative $\eps$, the two curves will always cross once
and only once, i.e., there is always a single positive root.
For positive $\eps$, the two curves may never cross, ``kiss'' once,
or cross twice, giving none, one and two positive roots, respectively.
There is a critical value of $a=\acrit$,
dependent on the particle masses and the positive value of $\eps$, 
at which there is a single positive root,
above which there are two, 
and below which there are none
(since $a$ is the exponent of a ratio less than one).
These results are illustrated in Fig.~\rf{schematic}.
%
%
\begin{figure}\centering
\includegraphics[width=9cm]{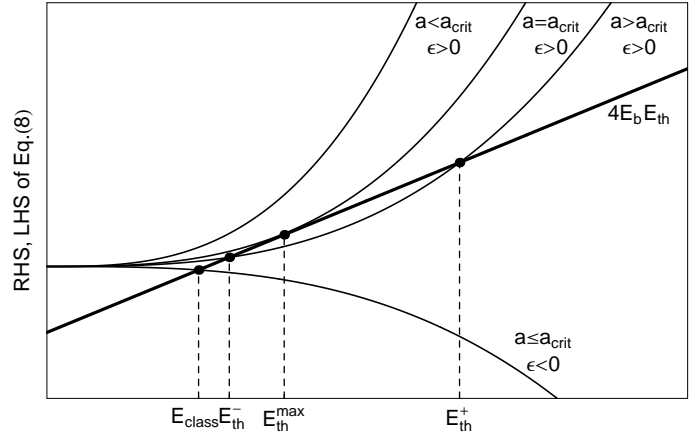}
\caption{Schematic illustration of the classical and modified threshold energies.}
\label{schematic}
\end{figure}
%

When $\eps <0$, the single positive solution for $\Eth$ is 
lower than the classical value $\Ecl$.  This leads to cutoffs
in the CR spectra at energies lower than those predicted from 
classical physics.
When $\eps>0$ and $a<\acrit$, there is no physical solution for $\Eth$,
and so the absorption reaction does not happen at any energy.
We return to this case briefly in \S\ref{sec:does}.
When $\eps>0$ and $a\ge \acrit$, then the solutions for $\Eth$
are always larger than $\Ecl$.
This leads to CR cutoff energies higher than those predicted from 
classical physics.  However, as $a$ increases, the influence of the 
foam term decreases, and for $a>\acrit$, 
the lower of the two solutions, 
call it $\Eminus$,  approaches the classical value $\Ecl$.
The higher of the two solutions, call it $\Eplus$, goes to $M_P$ as $a$ increases.

\section{Foam dynamics}
\label{sec:dynamics}
Not surprisingly, there exist several models for the foam dynamics
of $a$ and $\eps$.
We discuss some of these in this section.

\subsection{Foam Model with Fixed Fluctuation Parameter $\eps$}
\label{sec:fixedeps}
Since one interest of the particle-astrophysics 
community is to explain
possibly extended CR spectra, we first discuss the higher cutoffs 
provided by the $\Eth>\Ecl$ case, i.e., the 
$\eps>0$, $a\ge\acrit$ case.
If there were a reason in Nature to reject the lower $\Eminus$ 
solution (we know of none), then the arbitrarily-large 
value of $\Eth=\Eplus$ would allow CR cutoffs to be 
arbitrarily extended.
However, with both solutions operative, the reaction will occur when 
$E$ rises to exceed either solution, 
i.e.\ to exceed $\min\{\Eminus,\Eplus\}$,
which is just the lower-energy solution $\Eminus$;
the higher $\Eplus$ solution seems irrelevant.
So, how large can $\Eminus$ be?
The answer is that $\Eminus$ is maximized at the single solution value
occurring when $a=\acrit$.  Call this value $\Ethmax$.
These solutions and their labeling are shown in Fig.~\rf{schematic}.

The critical $a$ and the critical $\Eth$ can be found 
in principle by simultaneously solving two equations.\footnote{
In principle, Eq.~\rf{4} is sufficiently nonlinear that there could be
more than one solution for the critical set $\{ \Eth,\,a \}$.
However, more than a single solution is not apparent in our numerical work.
In our Appendix we explain why the single solution for $\acrit$ is effectively unique.
}
The first is just \rf{4},
and the second is obtained from \rf{4} by equating the 
first derivatives of the LHS and RHS with respect to $\Eth$.
However, manipulation of these two equations does not lead to 
a useful analytic separation of $\acrit$ and $\Ethmax$.
We content ourselves to use numerical techniques in the main text 
to determine $\acrit$ and $\Ethmax$,
but present some accurate analytical approximations in the Appendix.
There is however, one simple analytic relation that results from 
manipulations of these two equations.  It is
\beq{kiss}
\Ethmax=\Ecl\,\left(\frac{\acrit+2}{\acrit+1}\right)\,.
\eeq
This result shows that $\Ethmax$ depends on $\eps$ (assumed positive
here) only implicitly through $\acrit$, and that $\Ethmax$ lies in the 
interval $[\Ecl,2\,\Ecl ]$ regardless of the value of $\acrit$.
$\Ethmax$ approaches $2\,\Ecl$ as $\acrit\rightarrow 0^+$,
and approaches $\Ecl$ for $\acrit \gg 2$.
There are $\Eplus$ solutions exceeding $2\,\Ecl$,
but these are inevitably accompanied by a second solution, $\Eminus$, 
lying below $2\,\Eth$.
This is our first new result.
We repeat it:
{\sl For positive $\eps$, the reaction threshold energy can be {\bf raised},
but {\bf at most by a factor of 2}.}

Let us comment on the foam-inspired extended CR spectra obtained in
\cite{AC1,Ng1,Ng2}.
In this work a pre-desired value of $\Eth$ is input
into Eq.~\rf{4}, and from this the value of fixed, 
positive $\epsilon$ is extracted.
This approach suffers from (at least) two drawbacks.
The first is that it is oblivious to the existence of the second,
lower-energy solution $\Eminus$.
The second drawback is that it is fine-tuned in the value of $\eps$.
The first drawback is much more serious,
for we have just shown that the second solution
raises the threshold energy by at most a factor of 2.
The model of \cite{AC1,Ng1,Ng2} appears to fail to raise threshold energies.

\subsection{Foam Model with Gaussian Fluctuations}

A healthier approach to foam dynamics is described in \cite{Blasi}.
The parameter $\epsilon$ is treated as a stochastic variable,
subject to some specified probability distribution.\footnote
{The model of Ng et al.\ can be thought of as the special case 
where the stochastic distribution of $\eps$'s is sharply peaked about a 
pre-determined positive value and the lower of the two energy-thresholds
is ignored.
}
The stochastic assumption seems reasonable, in that 
space-time fluctuations at one site would not depend on the fluctuations elsewhere.
We follow this approach here, generalizing the results of \cite{Blasi}.

When an interaction occurs, the kinematics are determined by the
fixed parameter $a$ and a single random value of $\eps$.
Since experiments sum over many events,  
the total data sample is best described by the most probable value of the 
threshold energy.  This is determined by the 
Gaussian-distributed $\epsilon$.
In principle, some random occurrences will reduce the threshold 
for particular events even below the mean threshold.
However, our numerical work reveals that the width in $\Eth$,
resulting from the distribution in $\eps$, is small.  
This is evident in Figs.~\rf{gampackets}-\rf{cr23-1}.

In principle, the value of the fluctuation $\delta_{i}$ of each particle
can be treated as independent stochastic variables.
the effect of this on Eq.~\rf{4} is to replace the overall $\eps$
with an independent $\eps_a$, $\eps_c$, $\eps_d$ for the the 
respective three terms in the bracket.
In the end, an overall Gaussian is probably a good approximation
even for this more complicated case,
given the generality of the central limit theorem.
One might expect quantitative, but not qualitative, differences \cite{Blasi}.
We will treat the overall 
$\eps$ as the single stochastic variable, and define the probability distribution of 
the single $\eps$ as $p(\eps)$.
We will follow \cite{Blasi} and assume Gaussian statistics.
Then 
\beq{Gaussian}
p(\eps)= \frac{1}{\sigma\sqrt{2\,\pi}}\,
e^{-\frac{(\eps-\bar{\eps})^2}{2\,\sigma^2}}\,. 
\eeq
The stochastic $\eps$'s are then generated numerically via
\beq{epserf}
\epsilon=\sigma\sqrt{2}\;erf^{-1}(r)+\bar{\epsilon}\,\,,
\eeq
where 
$r$ is a random number in the interval $[-1,\,1]$,
$\sigma$ is the variance of a distribution, and $\bar{\epsilon}$ is
the average value of $\epsilon$. 
We set $\sigma=a$ to avoid introducing a new parameter.
Alternatively, one could for example 
choose a constant variance, say $\sigma=1$, or any other value.
This does not change the general behavior of our results,
as we show in the next section.
We choose $\bar{\epsilon}=0$ based on a preference for symmetry,
and to maintain the smallness of fluctuations.
The same choice was made in \cite{Blasi}.
A model with nonzero $\bar{\epsilon}$ was proposed in \cite{Gallou}. 
Obviously, this expresses a preference for 
negative $\eps$ (lowered threshold) over positive $\eps$ (raised threshold) 
or vice versa.  While this asymmetrical choice may turn out to merit Nature's
attention, it has not yet attracted our attention.

With the symmetrical choice for $\eps$,
half of the fluctuations present negative $\eps$, and half
present positive.  For the negative half,
each $\eps$ generates one solution for $\Eth$,
with $\Eth<\Ecl$.
For the positive half, 
each $\eps$ generates no solution when $a<\acrit$,
and two solutions above $\Ecl$ when $a>\acrit$.
The lower of these two solutions, $\Eminus$, is relevant,
while the higher solution is probably not.

\section{Modified Thresholds in Detail}
\label{sec:CRs}
For gamma-rays incident on the IRB,  Eq.~\rf{4} becomes
\beq{n1}
E_{IRB}\Eth=m_{e}^{2}+\eps\,\frac{\Eth^{2+a}}{M_{P}^{a}}\,\frac{2^{a}-1}{2^{2+a}}\,,
\eeq
where $m_e$ is the electron mass,
and for definiteness we take $E_{IRB}=0.025$~eV.
For CR nucleons interacting on the CMB, Eq.~\rf{4} becomes
\beq{n2}
4E_{CMB}\Eth=(m_{p}+m_{\pi})^{2}-m_{p}^{2}
+\epsilon\,\frac{\Eth^{2+a}}{M_{P}^{a}}
\left[1-\frac{m_{p}^{1+a}+m_{\pi}^{1+a}}{(m_{p}+m_{\pi})^{1+a}}\right]\,,
\eeq
with $m_p$ and $m_\pi$ the nucleon and pion masses.
Here, for definiteness we take $E_{CMB}=7.2\times 10^{-4}$~eV,
near the mean energy of the spectrum.
Solving these equations numerically, 
we map the random Gaussian distribution of 
$\eps$ described in Eq.~\rf{epserf} onto a random fluctuation spectrum for $\Eth$.

Different choices of $a$ characterize different foam models.
The two choices $a=\frac{2}{3}$ and 1 are motivated 
by interesting plausibility arguments \cite{Ng1,Ng2}.
Integral values of $a$ are motivated by loop quantum
gravity \cite{loop}, and also by arguments for unbroken rotational 
invariance \cite{Lehnert}.   
We take the agnostic approach and treat $a$ as a continuous parameter 
to be explored from zero upward. 
In Figs.~\rf{acritgam} and \rf{acritCR} we show 
the evolution of solutions with $a$.
Each ``solution'' $\Eth$ is really 
the most probable value of $\Eth$ picked from 
a distribution.\footnote
{Numerical work reveals that for $\eps<0$ ($\eps>0$),
the most-probable value for $\Eth$ is given by the solution
to Eqs.~\rf{n1}, \rf{n2} when $\eps$ is set roughly equal to $-\sigma$ 
for $\eps<0$,
and set to $+\sigma$ for $\eps>0$.
This is sensible,
as restricting the two sided distribution $p(\eps)$ to just one 
side moves mean $\eps$ from zero to $\pm\sqrt{{2}{\pi}}\,\sigma$.}
%
%
\begin{figure}\centering
\includegraphics[width=9cm]{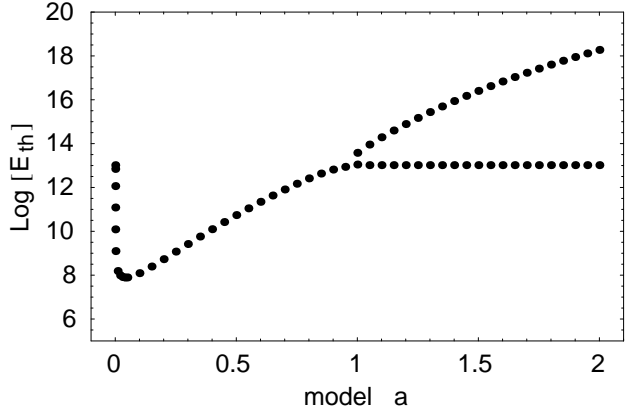}
\caption{Most probable values of threshold energies (in eV) 
vs. foam models for $\gamma$-rays}
\label{acritgam}
\end{figure}
\begin{figure}\centering
\includegraphics[width=9cm]{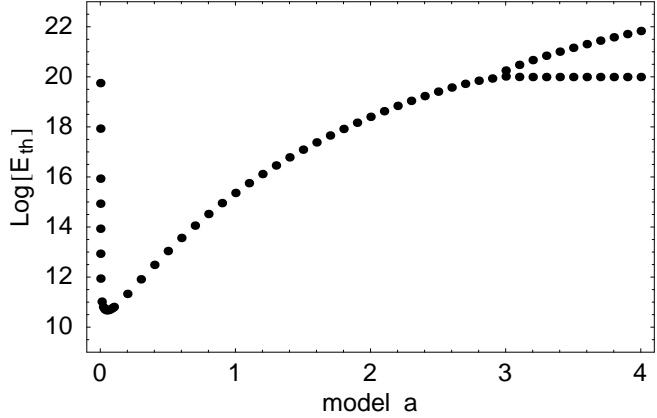}
\caption{Most probable values of threshold energies (in eV) 
vs. foam models for UHECRs.}
\label{acritCR}
\end{figure}
%

The values of $\acrit$ are evident in the critical points 
of these figures. Numerically, they are $\acrit=0.964$ for the gamma reaction,
and $\acrit=2.87$ for the nucleon reaction. 
Numerical values of $\acrit$ in general depend on the variance $\sigma$ 
in the $p(\eps)$ distribution, here set for simplicity to a. 
When $\sigma$ is taken as a free parameter, we find that the following 
limiting values for $\acrit$ would result:
$\acrit\rightarrow 0$ when $\sigma\rightarrow\infty$, 
and $\acrit\rightarrow\infty$ when $\sigma\rightarrow 0$.

Below $\acrit$, the single curve reveals the 
single solution accompanying negative $\eps$ fluctuations.
Above $\acrit$, and accepting both positive and negative $\eps$ fluctuations,
three solutions are actually present.
The two relevant solutions, $\Eth$ from negative $\eps$ fluctuations,
and $\Eminus$ from positive $\eps$ fluctuations, are nearly identical
in value, clustered just below and just above, respectively, 
the classical solution $\Ecl$.
In the figure, these two near-classical solutions 
constitute the unresolvable horizontal branch of the curve
to the right of the critical point, while the 
irrelevant solution $\Eplus$ is the curve that rises
to the right $\acrit$.

In Figs.~\rf{gampackets} and \rf{CRpackets}, 
the individual solution ``packets'' are shown, 
for different values of the parameter $a$.
%
%
\begin{figure}\centering
\includegraphics[width=9cm]{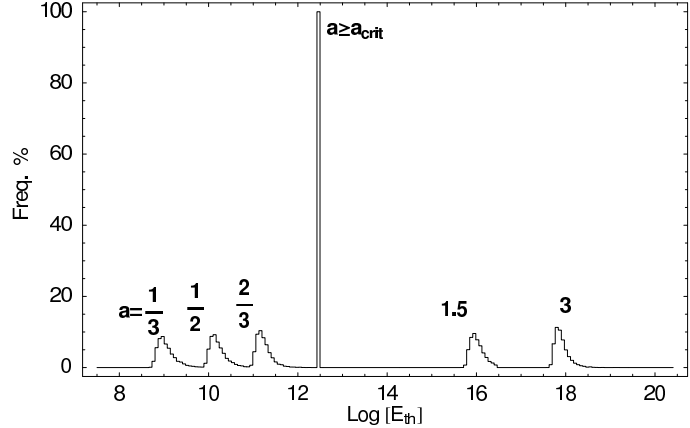}
\caption{Distributions of threshold solutions for gamma-rays}
\label{gampackets}
\end{figure}
\begin{figure}\centering
\includegraphics[width=9cm]{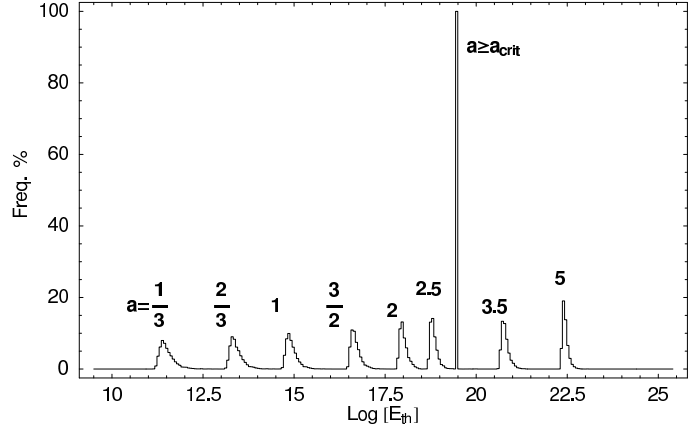}
\caption{Distributions of threshold solutions for UHECRs}
\label{CRpackets}
\end{figure}
%
%
The area of each packet reflects how often a fluctuation in 
$\eps$ produces a physical solution.
Below $\acrit$, the total area in the packet is 50\%,
since only half of the fluctuations, the $\eps<0$ ones,
produce a physical solution.
At $\acrit$, the sharp packet has total area of 100\%,
reflecting one solution for each $\eps<0$ and one for each $\eps>0$.
Above $\acrit$, 100\% of the area remains in the sharp peak
labeled $a\ge\acrit$, coming from the $\eps<0$ solution
and the $\Eminus$ solution, plus another 50\% area exists in
the higher-energy $\Eplus$ solution.
Raising $a$ reduces the magnitude of the space-time fluctuations,
and so pushes the packets comprised of the 
$\eps<0$ and $\Eminus$ solutions closer to $\Ecl$.
Raising $a$ also pushes the $\Eplus$ solution ever higher, 
toward $M_P$.

We note that the width of the $\eps<0$ packet 
decreases as $a\rightarrow \acrit^-$.
Above $\acrit$, the $\Eminus$ width and $\eps<0$ solution width  
remain narrow, the former related to the bounding of 
$\Eminus$ between $\Ecl$ and $2\,\Ecl$.
The $\Eplus$ width above $\acrit$ gets narrower as $a$ increases,
or equivalently, as $\Eplus\rightarrow M_P$, due to the 
$\frac{1}{M_{P}^{a}}$ suppression of the fluctuation.

In Figs.~\rf{gam23-1} and \rf{cr23-1} we show the $\Eth$ packets
for gamma-ray reactions and nucleon reactions, for $a=\frac{2}{3}$ and 
$a=1$.  These two values are among the most popular in the literature.
%
\begin{figure}\centering
\mbox{\includegraphics[width=7cm]{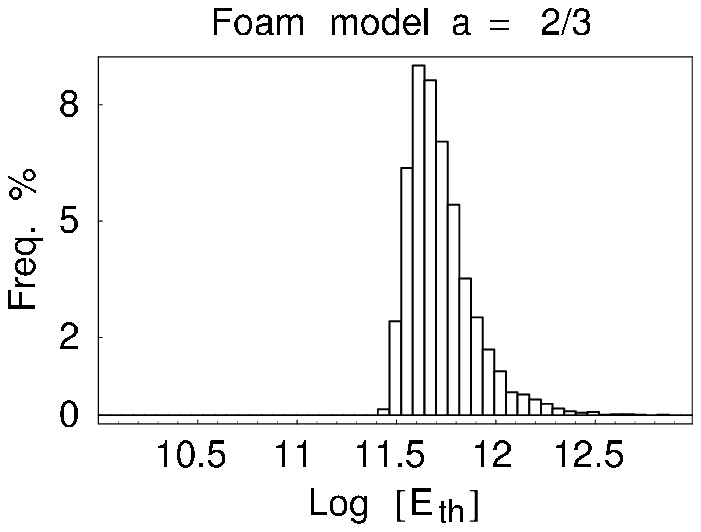}\quad\includegraphics[width=7cm]{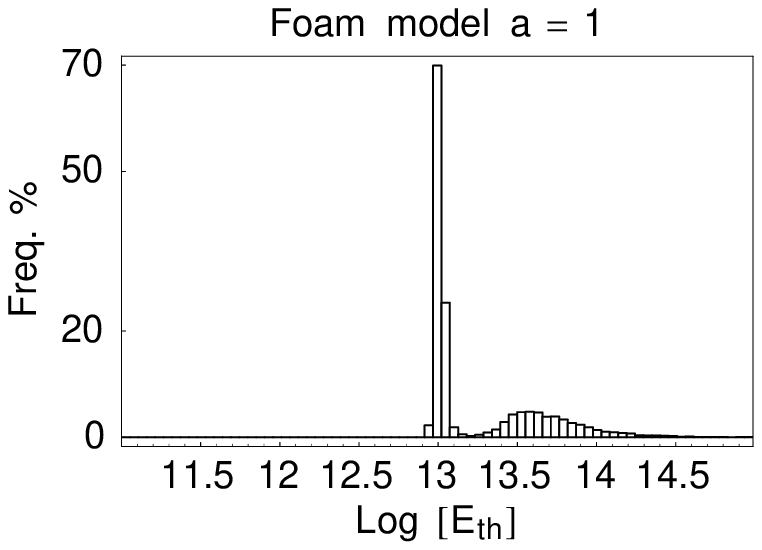}}
\caption{Threshold Distributions for $\gamma$ rays (energy in eV)}\label{gam23-1}
\end{figure}
\begin{figure}\centering
\mbox{\includegraphics[width=7cm]{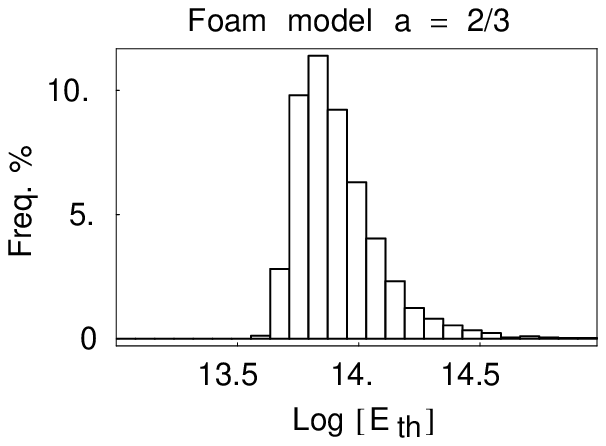}\quad\includegraphics[width=7cm]{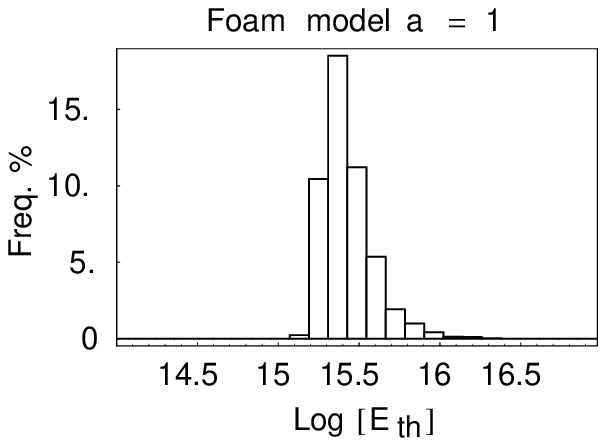}}
\caption{Threshold Distributions for UHECRs (energy in eV)}\label{cr23-1}
\end{figure}

Note that $a=\frac{2}{3}$ is below $\acrit$ for both 
nucleon and gamma-ray reactions.
Accordingly, there is a single solution for $\Eth$ 
from $\eps<0$ and none from $\eps>0$.
The $\eps<0$ solution lowers $\Eth$ below the classical threshold.
On the other hand, the $a=1$ value is again below $\acrit$ for 
the nucleon reaction, but above $\acrit$ for the gamma-ray reaction.
Thus, the $a=1$ gamma-ray distribution shows the nearly-classical
threshold, as well as the irrelevant higher-energy solution.

\section{An Asymmetric Foam Model that Does Extend the Spectra}
\label{sec:does}
We have shown that one can only lower the reaction thresholds
when fluctuations to negative $\eps$ are allowed.
We have also shown that positive $\eps$ fluctuations raise the 
threshold energy by at most a factor of two when $a>\acrit$,
but yield no physical solution for $\Eth$ when $a<\acrit$.
Thus, there is but a single way to raise the reaction thresholds:
the stochastic variable $\eps$ must be restricted to positive values, 
and $a$ must be below $\acrit$;
the absence of a physical threshold in this case means that
the absorption reaction cannot happen at all.

To postulate that $a<\acrit$ seems acceptable.
After all, the values for $a$ suggested by modelers are small,
as we have discussed.
The concomitant requirement that stochastic $\eps>0$,
i.e. $p(\eps)=0$ for all $\eps<0$,
seems harder to justify.
A Gaussian distribution cannot deliver this,
but any distribution of finite extent, e.g., a simple 
top-hat distribution or $\delta$-function distribution, can.
Perhaps Nature is this asymmetric, but single-sided distributions for 
fluctuations seem to us to be contrived,
and we honorably choose not to pursue them here.

\section{Discussion, with Speculations}
\label{sec:discuss}
In this section we present some additional interesting issues of the foam model.

\subsection{TeV-scale Gravity?}
The numerical results we have presented 
depend on the assumption that the Planck 
mass has its usual value $1.2\times 10^{28}$~eV.
Modern thinking, inspired by the extra dimensions available in string theory,
admits the possibility that the fundamental mass of gravity may be much 
less than this value.
One may ask what changes in our analysis with this (much) smaller mass-scale.
Qualitatively, nothing changes.  
The construction of solutions for positive or negative
$\eps$ values, illustrated in Fig.~1 is unchanged.
Also, the relation between $E^{\rm max}_{\rm th}$ and $E_{\rm class}$ 
given in Eq.~\rf{kiss} is independent of $M_P$, and so is unchanged.
So it is just the quantitative values of the new threshold energies that 
are changed.
But the quantitative changes can be dramatic.

%
%
%
In our numerical work we have investigated the effect of lowering 
$M_P$.  We find that, except for very small values of $a$,
$\Eth$ cannot exceed $M_P$.  This means that 
the Planck mass $M_P$ provides an energy limit for all cosmic rays.
Turning this remark around, the observation of cosmic rays above 
$10^{20}$~eV implies, in the context of this LIV model, that 
$M_P > 10^{20}$~eV.
The bound is eight orders of magnitude below the usual 4-dimensional 
Planck mass.
However, the bound is also incompatible with the popular TeV-scale 
gravity models by about eight orders of magnitude.

\subsection{Proton and Photon ``decays''}
The general form of Eq.~\rf{4} enables us to discuss other 
processes forbidden by exact Lorentz invariance.
Above some threshold energies, certain particle ``decays'' may 
become kinematically allowed \cite{Blasi}.
In \cite{Blasi,Gallou}, the authors discuss the two specific decays 
$p\rightarrow p\pi^{0}$ and $\gamma\rightarrow e^{-}e^{+}$
for the fixed value $a=1$.
In this section we will follow \cite{Blasi} and assume that energy-momentum fluctuations 
of the initial state particle characterize the Lorentz violation.
However, we generalize their results by
treating $a$ as a continuous free parameter.
Following a treatment similar to that which led to Eq.~\rf{4}, we here find
the energy-threshold for the process $a\rightarrow c+d$ to be
\beq{dec1}
E_{th}=\left\{\left[(m_{c}+m_{d})^{2}-m_{a}^2\right]
\frac{(M_{P})^{a}}{-\epsilon}\right\}^{1/(a+2)}\,.
\eeq
One may invert this equation to isolate the $a$-parameter:
\beq{dec2}
a=\frac{\log\left[\frac{1}{-\eps\Eth^2}\left[(m_{c}+m_{d})^{2}-m_{a}^2\right]\right]}
	{\log\left(\frac{E_{th}}{M_P}\right)}\;.
\eeq
It can be easily seen from the form of \rf{dec1} that only for $\epsilon<0$ 
is the threshold energy $E_{th}$ positive and real. 
For $\epsilon>0$ there is no physical solution.
Furthermore, for any negative $\eps$ and any $a$, 
there is a unique $E_{th}$.
We present our results in Fig.~\rf{fig:decays} for $-\eps= 1$.
\begin{figure}\centering
\includegraphics[width=9cm]{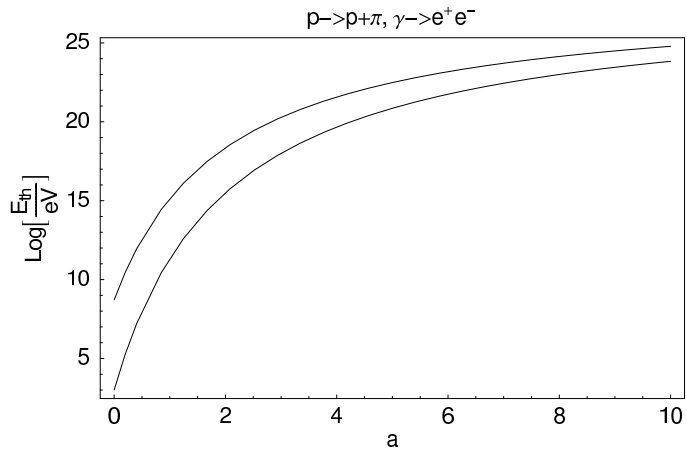}
\caption{Threshold energies for $p\rightarrow p\pi^{0}$ (higher curve) and $\gamma\rightarrow
e^{-}e^{+}$ (lower curve) decays as a function of parameter $a$.}
\label{fig:decays}
\end{figure}
As with the scattering processes presented earlier, 
here too predicted thresholds are bounded from below by some
simple function of particle masses (given in Eq.~\rf{dec1}), 
and from above by $M_{P}$.
For the $a=1$ case of \cite{Blasi}, 
one obtains $E_{th}\sim 10^{15}eV$ for ``pion-bremsstrahlung'' by a free proton, 
and $\Eth\sim 10^{13}eV$ for photon decay to $e^+ e^-$. 
To hide the particle decays, one may move the threshold energies beyond the highest 
observed cosmic ray energies ($\Ecl$) by raising the value of $a$.  
Doing this, one gets from Eq.~\rf{dec2} a lower limit for allowed values of $a$.  
To be specific, inputing $10^{20}$~eV for cosmic rays,
and $10^{13}$~eV for gamma-rays, one obtains for $\eps=-1$
the lower bounds of 
$2.82$ and $0.93$ for the respective $a$'s.\footnote
{We note that the lower limits for $a$'s from particle decays are 
very similar to the lower limits for $a$'s from the related scattering processes,
given by the $\acrit$ values obtained in \S\ref{sec:CRs}.
}

\subsection{Cosmic-Ray ``Knees''}
We note that although our results suggest failure for the 
program which attempts to {\sl raise} cosmic-ray absorption thresholds,
the lowered threshold may nonetheless have application.  
It is conceivable that structure in the cosmic-ray spectrum,
e.g., the first and second  ``knees'', and the ``ankle'',
are results of lowered absorption thresholds.
In particular, if the parameter $a$ itself were to trace 
the matter distribution, along the lines mentioned above for 
inhomogeneous dark energy, 
then $\Eth$ for galactic cosmic-rays could be much lower than 
for the extra-galactic component.  The net result might be spectral breaks
at these lower $\Eth$ values for cosmic-rays contained in our Galaxy, 
i.e., ``knees''.  The argument disfavoring this remark 
is that Galactic cosmic rays are probably not contained long enough 
to undergo absorption (assuming, as we have here,
that the absorption cross-section does not depend on $\eps$ or $a$).
%

\subsection{ A Random Walk Through Foam?}
There appear to us to be conceptual problems
associated with the kind of model studied here.  
For example, it is an assumption in the model that energy-momentum
is transferred between the particle and the foam 
at the point of particle interaction.
However, energy-momentum transfer between a free particle and
the foam must be disallowed, 
for this would present a random wall through Planck-sized domains,
leading to, e.g., unobserved angular deflections of light.
Resolution of these kinds of {\sl ad hoc} rules,
e.g., smoothing of the foam, must await 
a true theory of quantum gravity.
In advance of a theory of quantum gravity, 
speculations are allowable, and in the next subsection 
we provide a few.

\subsection{Energy-Momentum Non-conservation}
In foam models, energy and momentum are not generally conserved 
in particle interactions.
Non-conservation is suppressed for interactions at terrestrial 
accelerator energies and below, but possibly 
becomes noticeable for interactions of extreme-energy cosmic-rays
with cosmic background fields, or with atmospheric nuclei.
Obvious questions are how much energy-momentum is missing, 
and where does it go?
To the best of our knowledge,
these questions are not addressed in the literature.

In the context of the model analyzed in this work,
it is simple to estimate the energy-momentum loss of the interacting
quanta.  As shown in Eq.~\rf{2a}, the energy-momentum added 
to the quanta is $\delfoam$, which, obtained from Eq.~\rf{2b}, is
\beq{delfoam1}
\delfoam=4\,E_b\,\left(1-\frac{\Ecl}{\Eth}\right)\,.
\eeq
This result makes it clear that a lowered (raised) threshold,
$\delfoam<0$ ($\delfoam>0$), and some particle energy-momentum is lost 
(gained).

Since the numerical value of $\Eth$ depends on $a$, 
$\delfoam$ too depends on $a$.
We may see this directly by comparing Eqs.~\rf{2b} 
and \rf{4} and to deduce that 
\beq{delfoam2}
\delfoam= \eps\,\Eth\,\left(\frac{\Eth}{M_P}\right)^{a}\,
	\left[1-\frac{m_{c}^{1+a}+m_{d}^{1+a}}{(m_{c}+m_{d})^{1+a}}\right]\,.
\eeq
This equation also makes it explicit that the sign of $\delfoam$ 
is the same as the sign of $\eps$.
Thus, particle energy-momentum is lost (gained) when $\eps<0$
($\eps>0$).

We have seen that when $a<\acrit$, there is no solution 
for $\Eth$ for positive $\eps$, and a single solution, typically 
well below $\Ecl$, for negative $\eps$.
Thus, stochastic fluctuations will on average drain energy-momentum 
from the interacting particles, when $a<\acrit$.
For $a>\acrit$, $\Eth$ is very nearly $\Ecl$, and there is negligible
energy-momentum transfer.
An example of significant energy-momentum loss is displayed in
Fig.~\rf{fig:deltafoam}, where 
a distribution of $\delfoam$ versus negative $\eps$ is shown 
for cosmic-rays, with the sub-critical value $a=1$.
%
%
\begin{figure}\centering
\includegraphics[width=9cm]{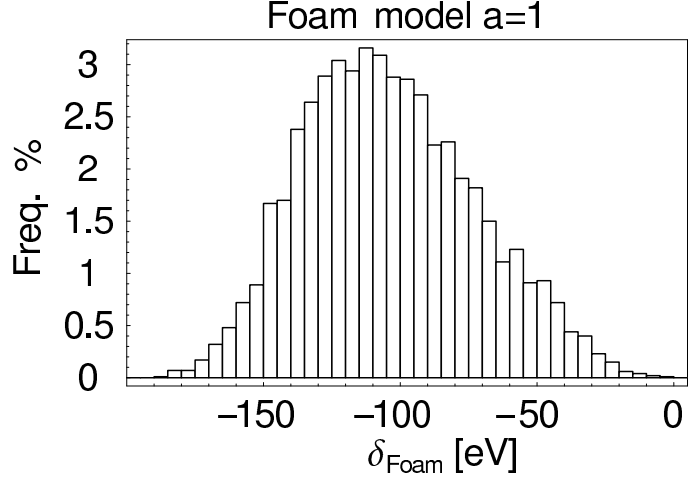}
\caption{Energy-momentum transferred to vacuum foam, as a function of
negative $\eps$ (the case where $\Eth$ is lowered),
for $a=1$.}
\label{fig:deltafoam}
\end{figure}
%
%
The typical missing energy-momentum per interaction is 
$\sim 120$~eV, consistent with Eq.~\rf{delfoam1}.
As a fraction of the interaction energy-momentum,
this loss is tiny.
From Eq.~\rf{delfoam1}, when $\Eth\ll\Ecl$ 
one has for the fractional loss,
$\delfoam/\Ecl = -4\,E_b/\Eth \sim 
10^{-18}(E_b/E_{\rm CMB})({\rm PeV}/\Eth)$, 
far too small to ever be detected by direct cosmic-ray measurements.

Where the missing energy-momentum goes is a difficult question.
Without a theory of quantum gravity, one is reduced to speculation.
Perhaps the vacuum of the Universe itself recoils in the
interaction (similar to Bragg diffraction, 
or the M\"ossbauer effect).\footnote
{In the related model of \cite{EllisDbrane}, energy-momentum is transferred
to the extra-dimensional bulk of string theory, 
via ``D-particle'' dynamics.
}
In an isotropic Universe, the net three-momentum transferred
will be zero, after averaging over many interactions.
So our Universe is not running away from us.
However, the net energy in the vacuum will grow with each interaction.
Perhaps an integration over the interaction history provides the 
observed ``dark energy'' of the Universe, with equation of state
$\frac{p}{\rho}=-1$.

Or perhaps the missing energy-momentum is not homogeneously distributed.
For example, cosmic ray interactions with galactic matter may deposit 
missing energy-momentum into the vacuum of galaxies.  Then one 
might expect dark energy to exhibit some ``clustering'' around 
large-scale matter distributions.  
Such a dark energy over-density might appear to be ''dark matter''.

An alternative point of view is that the missing particle energy 
is simply not missing.
For example, since $\delfoam\equiv\half\,(\delta E-\delta p)$,
perhaps $\delta E=0$ and only momentum is lost in the interaction.
The formalism in this paper goes through unchanged in such a case.
As for the momentum, even though momentum is lost from the particles
in each interaction, the isotropy of the Universe guarantees that 
after averaging over momenta directions from many interactions, 
no net momentum is gained or lost by the vacuum.
Thus, there may be no net transfer of energy-momentum to the vacuum.

Particle physics in the expanding Universe may provide a useful focus for thought.
In the expanding Universe, the momentum 
in a co-moving collection of particles is not conserved (it red-shifts),
even though the local Einstein equations 
conserve the energy-momentum of interacting particles.
This global non-conservation of energy is related to the lack 
of time-translation symmetry in the global, expanding Universe.
Perhaps in the foam models, translation symmetry is sufficiently broken, or 
the interaction of the particles with the metric foam is sufficiently 
nonlocal, to make the apparent energy-momentum non-conservation palatable.
Even the concept of local versus global becomes confused when the particle
interaction involves the foam.  If the foam fluctuation is included in the 
``local'' environment of the interaction, then energy-momentum 
can be said to be locally conserved.
However, if the foam is counted as part of the global vacuum,
then energy-momentum is not conserved locally,
and may or may not be conserved globally.

\subsection{Relation to Modified Dispersion Equation Approach}
Finally, we remark on the relation between modified interaction kinematics, 
as presented in this work, and related work on 
modified particle dispersion-relations.
It turns out that the same modified threshold equation 
is obtained in either approach, with one significant difference:
the relative sign of the stochastic variable $\eps$ 
is opposite in the two approaches \cite{AC1,Ng1}.
If one includes both 
modified kinematics and modified dispersion relations,
then the two foamy effects cancel each other.\footnote
{Of course, this assumes the same form for the fluctuations $\eps$
in both the kinematics and the dispersion relation.
}

\section{Conclusions}
We have analyzed a family of Lorentz-violating, space-time foam models.
Fluctuation amplitudes are assumed to be stochastic, 
Gaussian-distributed, and suppressed by the Planck mass.
These models can be tested, in principle, by searching 
for anomalous propagation of high-energy cosmic-rays and gamma-rays.
We have derived an equation for a new
(modified) threshold energy, in the general case of two particle scattering.
As relevant examples, we examined the modified energy-thresholds 
for the reactions $N+\gamma_{CMB}{\rightarrow}\Delta\rightarrow N'+\pi$
and $\gamma+\gamma_{IRB}{\rightarrow}e^{+}+e^{-}$
affecting propagation of extreme-energy cosmic rays 
and TeV gamma-rays, respectively.
Our threshold solutions can be parametrized by a suppression parameter $a$. 
For a given particle reaction, there exists a critical value $\acrit$ 
of this parameter, beyond which foam does not alter the standard predictions.
Furthermore, for $a<\acrit$, foam does alter the reaction
significantly, but always lowering, never raising, interaction thresholds.
We found that $\acrit\sim 3$ for the 
$N+\gamma_{CMB}{\rightarrow}\Delta\rightarrow N'+\pi$ reaction,
and $\acrit\sim 1$ for the 
$\gamma+\gamma_{IRB}{\rightarrow}e^{+}+e^{-}$ reaction.

Previously, Aloisio et al.\ \cite{Blasi} investigated one model 
(corresponding to our $a=1$ case) for its effect on the 
UHECR absorption reaction on the CMB.
They found that the reaction threshold is lowered to a 
most probable value of $\Eth\simeq 2.5\times 10^{15}$~eV. 
Since the $a=1$ used in \cite{Blasi} is below $\acrit$, their lowered threshold 
is consistent with our more general results
which extend their model to arbitrary values of $a$. 
Even for $a>\acrit$, we found that the threshold can be raised only by a factor of at 
most two (according to Eq.\rf{kiss}), and even then only if 
fluctuations are asymmetrically distributed about zero.
Thus, it appears that spacetime foam models of the kind we assessed
are unable to extend the spectra of UHECRs or gamma-rays beyond their 
classical absorption thresholds on the CMB and IR background, respectively.

Tests of foam models by means other than anomalous cosmic-ray propagation
abound in the literature.
These include, but are not limited to,
proton and even photon decay, neutron stabilization,
anomalous UHECR shower development, \ldots .
Although a discussion of all these possible signatures is beyond the 
focus of this work, we did investigate the energy-thresholds above which 
kinematically disallowed $1\rightarrow 2$ processes become allowed.
The thresholds for these processes increase monotonically with the $a$-parameter.
For the processes 
$N+\gamma_{CMB}{\rightarrow}\Delta\rightarrow N'+\pi$
and 
$\gamma+\gamma_{IRB}{\rightarrow}e^{+}+e^{-}$,
the energy-thresholds can be pushed beyond the observed end-points 
of cosmic-ray and gamma-ray spectra with $a$ values (remarkably) similar 
to respective $\acrit$ values.

We have succumbed to temptation and presented a few speculations.
One is that the transfer of energy between interacting particles 
and foam may contribute to dark energy, or to apparent 
inhomogeneous dark matter.
Another is that lowered thresholds in cosmic ray absorption may effect the 
``knees'' (increase in the spectral slope) observed in the Galactic 
cosmic-ray spectrum.

Finally, we noted that the absorption thresholds for 
cosmic-rays and gamma-rays, in the context of the model, cannot exceed the 
``Planck mass''.  Thus, the observation of cosmic rays with energies at 
$10^{20}$~eV necessitate a fundamental Planck mass in excess of $10^8$~TeV.
This model is, then, incompatible with TeV-scale gravity models, 
by eight orders of magnitude.

\vspace{1.0cm}

\noindent
Acknowledgments: 
This work was supported in part by DOE grant DE-FG05-85ER40226.

\begin{appendix}
\section*{Appendix: Approximate Analytical Expressions}
In this appendix we derive approximate analytical expressions for
various quantities obtained in the paper by numerical methods. Only
the lowest order approximations will be given; it is straightforward
to calculate higher order corrections.

We start by rewriting Eq.~(\ref{4}) in dimensionless form:
\begin{equation}
x=1+\epsilon bc^{-a-1}f(a)x^{a+2},\label{eq:x}
\end{equation}
where
\begin{eqnarray}
&& x=\frac{E_{\text{th}}}{E_{\text{cl}}},\ \ \
b=\frac{M_{\text{Pl}}}{4E_b},\ \ \
c=\frac{M_{\text{Pl}}}{4E_{\text{cl}}},\ \ \ z=\frac{m_c}{m_d},\\ &&
f(a)=1-\frac{1+z^{a+1}}{(1+z)^{a+1}}.
\end{eqnarray} 
For the photon ($\gamma$) reaction 
$\gamma+\gamma_{IRB}{\rightarrow}e^{+}+e^{-}$
and nucleon ($n$) reaction 
$N+\gamma_{CMB}{\rightarrow}\Delta\rightarrow N'+\pi$,
the parameters are the following:
\begin{eqnarray}
b_\gamma\approx 1.22\times 10^{29}, & c_\gamma\approx 1.17\times
10^{15}, & z_\gamma=1,\\ b_n\approx 4.24\times 10^{30}, & c_n\approx
1.25\times 10^{8}, & z_n\approx 0.149.
\end{eqnarray}

We consider three different cases with fixed $\eps$: 
(1) positive $\epsilon$ and critical $a$; 
(2) positive $\epsilon$ and non-critical $a$; 
(3) negative $\epsilon$. 
Then we generalize to a distribution in $\eps$.  
Because of the plethora of subscripts and superscripts required in this appendix, 
we find it convenient to replace the subscript ``crit'' with the compact subscript ``$*$''.

\subsection{\bm{$\epsilon>0$},\ \ \ \bm{$a=\acrit\equiv a_*$}}

To determine the critical value of $a$, 
we need to solve Eq.~(\ref{eq:x}) and its derivative equation simultaneously.
These two equations are:
\begin{eqnarray}
&&x_*=1+\epsilon bc^{-a_*-1}f(a_*)x_*^{a_*+2},\label{eq:1:x}\\ &&1=\epsilon
bc^{-a_*-1}f(a_*)(a_*+2)x_*^{a_*+1}.\label{eq:1:dx}
\end{eqnarray}
Eliminating $x_*$ from Eqs.~(\ref{eq:1:x}) and (\ref{eq:1:dx}), we
find
\begin{eqnarray}
&& x_*=\frac{a_*+2}{a_*+1},\label{1:x}\\ && \epsilon
bc^{-a_*-1}f(a_*)=\frac{(a_*+1)^{a_*+1}}{(a_*+2)^{a_*+2}}.\label{1:a}
\end{eqnarray}
Eq.~(\ref{1:a}) cannot be solved exactly for $a_*$. It is, however,
very easy to find simple and accurate approximate solutions. To find
these, note first that as $a_*$ increases from zero to $\infty$, the
right-hand side of Eq.~(\ref{1:a}) monotonically decreases from
$\frac{1}{4}$ to zero as $(a_*e)^{-1}$, and the left-hand side first
increases from zero to its maximum and then exponentially decreases back to 
zero. Thus, in general, there can be none, one, or two solutions to
Eq.~(\ref{1:a}). For large $b$ and $c$ (as in both $\gamma$ and $n$
cases) and moderate $\epsilon$, there are two solutions: $a_{*0}\ll 1$
and $a_{*1}\sim 1$.

To find the smaller root $a_{*0}$, 
we expand terms in Eq.~(\ref{1:a}) to the
lowest order in $a_*$ and obtain
\begin{equation}
\epsilon a_{*0}\approx\frac{c}{4b}
\left[\ln{(1+z)}-\frac{z}{1+z}\ln{z}\right]^{-1}.
\end{equation}
Substitution of numerical values gives $\epsilon a_{*0\gamma}\approx
3.45\times 10^{-15}$ and $\epsilon a_{*0n}\approx 1.90\times 10^{-23}$.
Although these small values justify the approximation, they are too
small to be physically interesting for moderate values of $\epsilon$.
Thus we learn that for the relevant situation of large $b$ and $c$,
there is effectively a unique value for $\acrit$.

To find the physically interesting $\acrit\equiv a_{*1}$, 
we note that for moderate $a_*$ and large
$c$, the term $c^{-a_*-1}$ is the fastest changing part of
Eq.~(\ref{1:a}). Evaluating both the right-hand side of
Eq.~(\ref{1:a}) and the function $f(a_*)$ to be of order unity, we
arrive at the approximation $a_{*1}\approx a^{(0)}_{*1}+a^{(1)}_{*1}$,
where
\begin{eqnarray}
&&a^{(0)}_{*1}=-1+\frac{\ln{(\epsilon b)}}{\ln{c}},\label{astar1:0}\\
&&a^{(1)}_{*1}=\frac{1}{\ln{c}}\ln{\frac{\left(a^{(0)}_{*1}+2\right)^{a^{(0)}_{*1}+2}
f\left(a^{(0)}_{*1}\right)}
{\left(a^{(0)}_{*1}+1\right)^{a^{(0)}_{*1}+1}}}.\label{astar1:1}
\end{eqnarray}
Substituting numerical values, we find
\begin{eqnarray}
&&a^{(0)}_{*1\gamma}\approx 0.93+0.028\ln{\epsilon},\label{astar1:g}\\
&&a^{(0)}_{*1n}\approx 2.78+0.054\ln{\epsilon}.\label{astar1:n}
\end{eqnarray}

\begin{figure}[hbt]
\includegraphics[width=8cm]{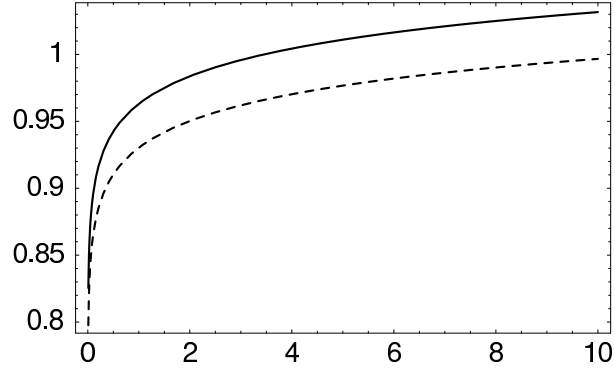}
\caption{Critical value $a_{*1}$ as a function of $\epsilon$ for the reaction
$\gamma+\gamma_{IRB}{\rightarrow}e^{+}+e^{-}$.  The dots 
represent the numerical solution, the dashed curve is the lowest order
approximation given by Eq.~(\ref{astar1:0}), and the solid curve includes 
the next order approximation given in Eq.~(\ref{astar1:1}).}
\label{fig:astar1-g}
\end{figure}

\begin{figure}[hbt]
\includegraphics[width=8cm]{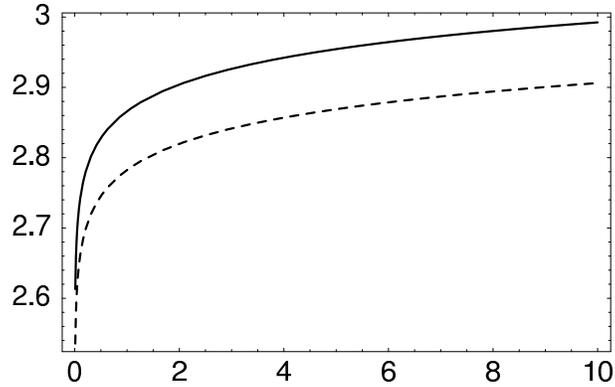}
\caption{As in Fig.~(\ref{fig:astar1-g}), but for the reaction
$N+\gamma_{CMB}{\rightarrow}\Delta\rightarrow N'+\pi$}
%
\label{fig:astar1-n}
\end{figure}

Figs.~\ref{fig:astar1-g} and \ref{fig:astar1-n} compare the lowest and
the next order approximations for $a_{*1}$ as a function of $\epsilon$,
with the corresponding numerical solutions for the $\gamma$ and
$n$ reactions, respectively.

\subsection{\bm{$\epsilon>0$},\ \ \ \bm{$a>\acrit\equiv a_{*1}$}}

As a function of $x$, Eq.~(\ref{eq:x}) with $a>a_{*1}$ and 
positive $\eps$ has two roots, $x_1$ and $x_2$. 
The term $\epsilon bc^{-a-1}f(a)$ in Eq.~(\ref{eq:x}) 
is of order unity for $a=a_{*1}$
and rapidly decreases for $a>a_{*1}$. Similarly, $\left[\epsilon
bc^{-a-1}f(a)\right]^{-1/(a+1)}$ is of order unity for $a=a_{1*}$ and
rapidly increases for $a>a_{*1}$. Thus for $a$ sufficiently above
$a_{1*}$, one root ($x_1$) is very close to unity and 
the other root ($x_2$) is very
large. Expanding Eq.~(\ref{eq:x}) correspondingly, we find the
following approximations for the two roots:
\begin{eqnarray}
&&
x_1\approx\frac{1-(a+1)\,\epsilon\,bc^{-a-1}f(a)}
{1-(a+2)\,\epsilon\,bc^{-a-1}f(a)}\,,
\label{x1}\\ 
\vspace{-0.2cm} & \nonumber\\
&&
x_2\approx \left[\,\epsilon\,
bc^{-a-1}f(a)\right]^{-1/(a+1)}\,.
\label{x2}
\end{eqnarray}

\subsection{\bm{$\epsilon<0$}}

For $\eps$ negative, Eq.~(\ref{eq:x}) has one root, $x_0$. 
Using the same arguments as in
the previous subsection, we conclude that for $a$ sufficiently below
$a_{*1}$, the root $x_0$ is very small; and for $a$ sufficiently above
$a_{1*}$, the root $x_0$ is very close to unity. Corresponding
expansions give
\beq{x0}
x_0\approx \left\{
\begin{array}{ll}
\left[\,|\epsilon|\,bc^{-a-1}f(a)\,\right]^{-1/(a+2)}\,; 
&
a<a_{*1}\,,
\\
\vspace{-0.2cm} & \\
\frac{1+(a+1)\,|\epsilon|\;bc^{-a-1}f(a)}{1+(a+2)\,|\epsilon|\;bc^{-a-1}f(a)}\,, 
& 
a>a_{*1}\,.
\end{array}
\right.
\eeq

\begin{figure}[hbt]
\includegraphics[width=8cm]{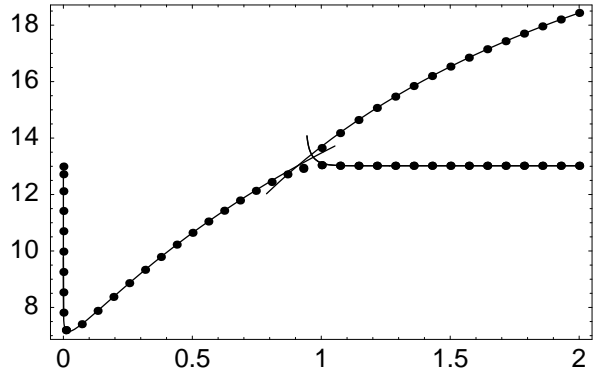}
\caption{Threshold energy in $\log_{10}{(E_\text{th}/\text{eV})}$ as a
function of $a$, for the reaction
$\gamma+\gamma_{IRB}{\rightarrow}e^{+}+e^{-}$.
The dots represent the numerical solution, and the curves
are approximations given by Eqs.~(\ref{x1}), (\ref{x2}), 
and (\ref{x0}) ($x_0$ above $\acrit$ and $x_1$ are indistinguishable on the plot). 
The curves are drawn slightly beyond their applicability ranges for easy
identification. The parameter $\epsilon$ is set to $\pm 1$.}
\label{fig:x-a-g}
\end{figure}

\begin{figure}[hbt]
\includegraphics[width=8cm]{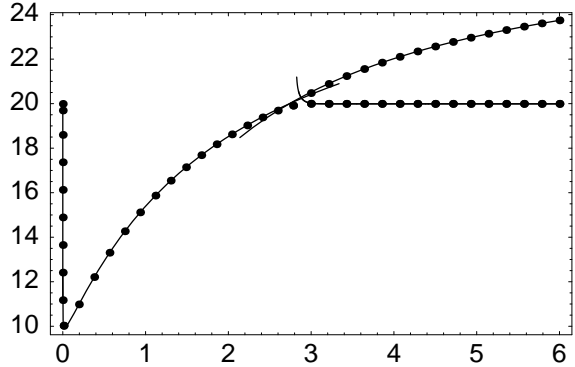}
\caption{As in Fig.~(\ref{fig:x-a-g}), but for the reaction
$N+\gamma_{CMB}{\rightarrow}\Delta\rightarrow N'+\pi$}
\label{fig:x-a-n}
\end{figure}

Figs.~\ref{fig:x-a-g} and \ref{fig:x-a-n} compare the lowest order
approximations [Eqs.~(\ref{x1}), (\ref{x2}) and
(\ref{x0})] for the roots as a function of $a$ with the corresponding
numerical solutions, for the $\gamma$ and $n$ reactions, respectively.

\subsection*{Stochastic Distributions}

The results of the previous subsections for fixed $\eps$ can be 
folded into a distribution in $\eps$ in a straightforward manner.
When the variable $\epsilon$ is distributed according to the
probability density function $p(\epsilon)$, the corresponding
probability density function $p(x)$ for the variable $x=x(\epsilon)$
is found from
\begin{equation}
p(x)=p(\epsilon)\left|\frac{d\epsilon}{dx}\right|.
\end{equation}
Two of our four solutions, $x_0$ with $a>\acrit$ in Eq.~\rf{x0} and 
$x_1$ in Eq.~\rf{x1}, are nearly independent of $\eps$.
So for these two solutions, whether $\eps$ is fixed or distributed is 
barely relevant.
However, the other two solutions, $x_0$ with $a<\acrit$ in Eq.~\rf{x0} and
$x_2$ in Eq.~\rf{x2}, are quite dependent on $\eps$.
Taking a normalized Gaussian distribution 
\begin{equation}
p(\epsilon)=(2\pi\sigma^2)^{-\frac{1}{2}}\exp{(-\epsilon^2/2\sigma^2)}
\end{equation}
as in the main text, 
and considering the roots $x_0$ for $a<a_{*1}$ and $x_2$ for $a>a_{*1}$,
we find
\begin{eqnarray}
p(x)\approx\left\{
\begin{array}{ll}
\displaystyle{(2\pi\sigma^2)^{-\frac{1}{2}}
\frac{(a+2)c^{a+1}}{bf(a)x^{a+3}} \exp{\left[-\frac{c^{2a+2}}
{2\sigma^2 b^2 f^2(a)x^{2a+4}}\right]}}, & a<a_{*1}\,;\\
\vspace{-0.2cm}& \\ 
\displaystyle{(2\pi\sigma^2)^{-\frac{1}{2}}
\frac{(a+1)c^{a+1}}{bf(a)x^{a+2}} \exp{\left[-\frac{c^{2a+2}}
{2\sigma^2 b^2 f^2(a)x^{2a+2}}\right]}}, &
a>a_{*1}\,.
\end{array}\right.
\label{p}
\end{eqnarray}

A few properties of the distribution $p(x)$ are evident from
Eq.~(\ref{p}): (i) $p(x)$ is asymmetric since it decreases
exponentially (as a power) to the left (right) of its peak; 
(ii) in the two intervals, $a<a_{*1}$ and $a>a_{*1}$, the
location and the height of the peak of $p(x)$ increases with $a$; 
(iii) in both intervals, the width of the distribution decreases with $a$. 
These three trends are evident in the figures of this paper.
To better compare $p(x)$ with its numerical counterpart
obtained in the paper, one would need to transform $p(x)$ into
$p(\log_{10}{x})=(\ln{10})\,x\,p(x)$ and adjust the normalization coefficient.
We stop short of this.
\end{appendix}


%

\end{document}